# Calibration Technique for Rotating PCB Coil Magnetic Field Sensors

Joseph DiMarco[1], Giordana Severino[2],

and Pasquale Arpaia[3]

1 FERMILAB, Batavia, Illinois, United States of America, Email:dimarco@fnal.gov
2 University of Sannio, Benevento, Italy
3 University Federico II, Naples, Italy, Email:pasquale.arpaia@unina.it
2/27/2019

**Abstract**

A high-accuracy calibration of inductive coil sensors based on Printed Circuit Board (PCB), commonly used in rotating coil field measurements of particle accelerator magnets, is presented. The amplitude and phase of signals with and without main field suppression are compared in order to simultaneously determine both the PCB rotation radius and the transverse offset of its plane from rotation center. The accuracy of planar wire placement on the PCB boards is exploited to create loops highly precise in area which rotate at different radii. Such an area reproducibility and circuit geometry allow the suppression of the fundamental field, enabling the calibration, as well as improving signal resolution and mitigating vibration effects. Furthermore, the calibration can be performed dynamically, in-situ during measurements. Calibration accuracy is validated experimentally by referencing the PCB positions with a Coordinate Measuring Machine (CMM).

# 1. Introduction

Steering and focusing magnets are fundamental components of particle accelerators. The field quality of these magnets is influenced by several factors including (i) errors in construction geometry, (ii) iron saturation, (iii) coil deformation under electromagnetic forces, and (iv) eddy currents. Furthermore, cable effects, such as persistent currents, must be considered too in superconducting magnets [1]. Accurate field measurements are necessary to determine the harmonic content of a magnetic field, and compare it both to the accelerator requirements and to the results of simulations conducted during the design phase [2]. Rotating coils are the most accurate and widely used method to accomplish this task [3]. These probes consist of passive pick-up loops, measuring the voltage induced due to their rotational probe motion.

Assuming an ideal rotational motion of the probe, the accuracy in measuring the harmonic field components is determined by the uncertainty of the probe winding geometry. Calibration of these geometries about the axis of probe rotation is therefore central in obtaining accurate results [3]. For wire-based rotating coil sensors, e.g. where pick-up coils are implemented with multi-turn windings or Litz wire hand-wound on a grooved ceramic shaft [4][5], several standard methods have been developed to achieve good calibration [4-7]. Some of these require the use of high-accuracy linear stages, calibration magnets and special tooling [6][7]. In actual measurements, imperfections in motion of the probe [8][9], e.g. from vibrations, create additional error sources, not necessarily removed by the calibration process. However, suppression of the main field components by complementary windings, commonly referred to as "bucking" or "compensation", mitigates the spurious harmonic effects caused by the probe non-ideal motion [3][10][11]. Therefore, in addition to the calibration requirement, probes must have good bucking of the main fields to achieve high accuracy results.

Recent efforts in probe design have tried to take advantage of the precision achievable with PCB technology [14] and to enhance both the quality of bucking as well as the ease and accuracy of the calibration process [15]. This has been outlined previously [8], but without an explicit theoretical analysis as well as any experimental verification of the technique.

In this paper, this dynamic calibration technique for PCB coil magnetic field sensors is presented in detail. The main principle is that for inductive pick-up probes constructed using PCBs, the accuracy of planar wire placement - at least an order of magnitude better than that of careful machining - allows creation of loops nearly identical in area, but separated radially during rotation of the board as incorporated into a probe [16]. The precise areas give rise to bucking ratios on the order of 1000 (i.e. only 0.1% of the fundamental field remains), providing strong reduction of vibration effects. Moreover, the precision enables a complete, in-situ, and even rotation-by-rotation, probe calibration at the micron level: simultaneously determining the radius at which the PCB is rotating as well as the transverse offset of its plane from rotation center. To validate the technique and verify its accuracy, the experimental calibration results are cross-checked with Coordinate Measuring Machine (CMM) determinations of PCB displacements. In the following, Section 2 briefly recalls the theory behind rotating coils, Section 3 reports some considerations about the PCB design and manufacturing, Section 4 discusses the rotating coil sensor calibration, and Section 5 reports calibration results from an experimental test campaign. Conclusions follow in Section 6.

## 2. Rotating coil theory

Following the development in [3], the magnetic field within the aperture of an accelerator magnet can be expressed in terms of harmonic coefficients defined in a series expansion using a complex formalism:

$$B_y(z) + iB_x(z) = \sum_{n=1}^{\infty}(B_n + iA_n)\left(\frac{x + iy}{R}\right)^{n-1}, \tag{1}$$

where $z = x + iy$, $B_x$ and $B_y$ are the horizontal and vertical field components in the Cartesian coordinate system, and $B_n$ and $A_n$ are the $2n$-pole normal and skew harmonic coefficients at the reference radius $R$.

Changes in flux as the probe rotates are measured at the angular intervals of an optical rotary encoder and recorded with a digital integrator. Though the geometry of the pick-up loops on the probe can in principle be complicated, the measured flux change as a function of angle, θ, can be written straightforwardly as

$$\phi(\theta) = \text{Re}\left\{\sum_{n=1}^{\infty} C_n K_n e^{in\theta}\right\}, \tag{2}$$

where "Re" indicates the real part of the complex quantity, the field is defined by

$$C_n = B_n + iA_n, \tag{3}$$

and the complex probe sensitivity, $K_n$, is the sum over all wires on the probe

$$K_n = \sum_{j=1}^{N_{wires}} \frac{L_j R}{n}\left(\frac{x_j + iy_j}{R}\right)^n (-1)^j \tag{4}$$

(here $L$ is the length of a given wire and it can be set to 1 if the probe is longer than the magnet). The $(-1)^j$ gives the sign of the current flow of each wire and the $(x_j, y_j)$ are the locations of the wires with respect to the rotation axis. The harmonic fields can then be determined by obtaining the complex Fourier coefficients $F_n$ from an FFT of the measured $\phi(\theta)$ data, and then dividing by the complex sensitivities. That is

$$C_n = \frac{F_n}{K_n}. \tag{5}$$

The goal of calibration for any rotating coil is to accurately determine the positions of the wires with respect to the probe rotation axis so that the actual $K_n$ of the probe can be correctly determined. PCB-type probes have characteristics which allow for a particularly straightforward and convenient calibration of wire positions and therefore accurate determination of $K_n$. The PCB probes discussed here are of radial design. That is, with the rotating coil probe viewed in cross-section, the PCB lies in a radial orientation with respect to the probe rotation axis, with the etched loops placed parallel to each other at varying radial distances from the rotation axis (O) and extending along what would be the Z-axis (out of the page) as shown schematically in Fig. 1.

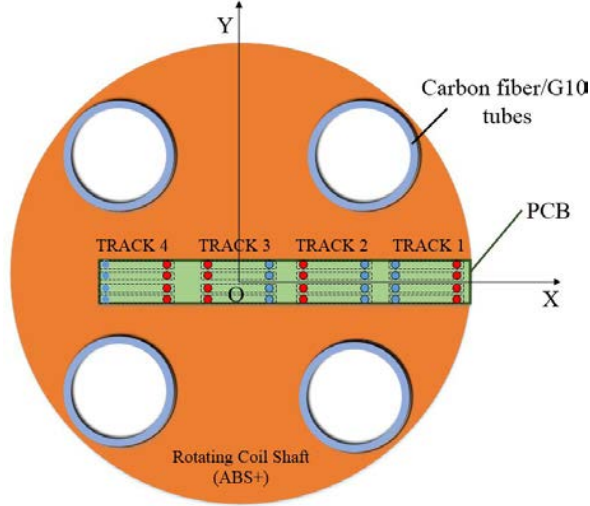

Figure 1: Cross-section of a PCB-based transducer of radial design (ABS+: Acrylonitrile butadiene styrene).

Since PCB manufacturing techniques locate traces at the level of micrometers on the surface in which they are etched (Fig. 2), both the parallelism and spacing of wires will be very well determined relative to other wires on the same PCB layer. Having a fixed and highly accurate *relative* location of the windings simplifies the calibration problem significantly, because, in general, there is no interest in determining the position of the individual wires, but only the absolute position of the ensemble - which if misaligned can have either horizontal (radial) offset, or vertical offset with respect to the rotation center. The calibration technique that determines these offsets is discussed in Section 4.2.

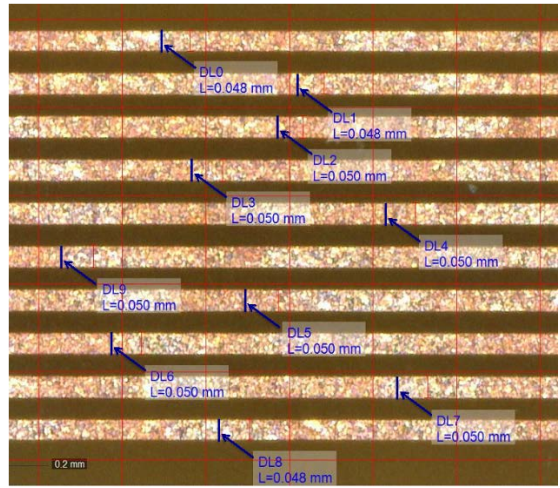

Figure 2: View on a PCB layer of one-side of the sensing coil turns (L: copper trace width measured by a calibrated microscope, DLi: label of the i-th measurement). Error and repeatability in the microscope readings is at the 1-2 μm level, though along the length of the board there may be variations in trace width of 5 μm on the 50 μm traces.

With regards to errors in measured harmonics caused by non-ideal probe motion, namely transverse and/or torsional vibrations during probe rotation, it has been shown in [3] that the most significant contribution to spurious harmonics comes from the fundamental field of the magnet, namely dipole or quadrupole components, typically larger than the higher order harmonics of eq. (1) by a factor of 1000-10000. Therefore it is desireable to measure the higher order harmonics using a probe winding which is insensitive to the dominant field distribution and to the harmonic order below it [3]. A simple way to achieve this with a probe of radial type is to have identical loops at different radial positions. For example, by considering a simple wire loop of width "w", with inner radius of the loop at position $x_{1b}$ (cross section view shown in Fig. 3), the dipole and quadrupole sensitivites from (2) will be

$$K_1 = LR\frac{x_{1b}+w}{R} - LR\frac{x_{1b}}{R} = Lw \qquad (6a)$$

and

$$K_2 = \frac{LR}{2}\frac{(x_{1b}+w)^2}{R^2} - \frac{LR}{2}\frac{x_{1b}^2}{R^2} = \frac{L}{2R}(2x_{1b}w + w^2) \qquad (6b)$$

To obtain a winding with no sensitivity to dipole, the loop described with Eqs. (6a), (6b) can be combined in series opposition with an identical loop having inner radius at position $x_{3b} = x_{1b} - d$ (Fig. 3). The combined sensitivity would then be

$$K_1 = Lw - Lw = 0 \qquad (7a)$$

and

$$K_2 = \frac{L}{2R}(2x_{1b}w + w^2) - \frac{L}{2R}(2(x_{1b}-d)w + w^2) = \frac{L}{R}(dw) \qquad (7b)$$

Note that equation (7b) is independent of the radial position, r, of the PCB. Since there is no sensitivity to dipole field (i.e. $K_1 = 0$), the dipole has been "bucked" from this winding. To achieve quadrupole as well as dipole bucking, an identical pair of windings, shifted radially with respect to those of Eqs. (7a), (7b) (loops 2 and 4 in Fig. 3), can be combined in series opposition to them. The shifted pair would again have zero dipole sensitivity and a quadrupole sensitivity as in eq. (7b). By putting them in series opposition with the previous two, it would leave $K_1 = K_2 = 0$. The outermost loop (loop1), used by itself, would then constitute an "unbucked" winding (UB), the series combination of loops 1–3 a "dipole bucked" (DB) winding, and the combination of loops 1-3+4-2 a "dipole-quadrupole bucked" (DQB) winding.

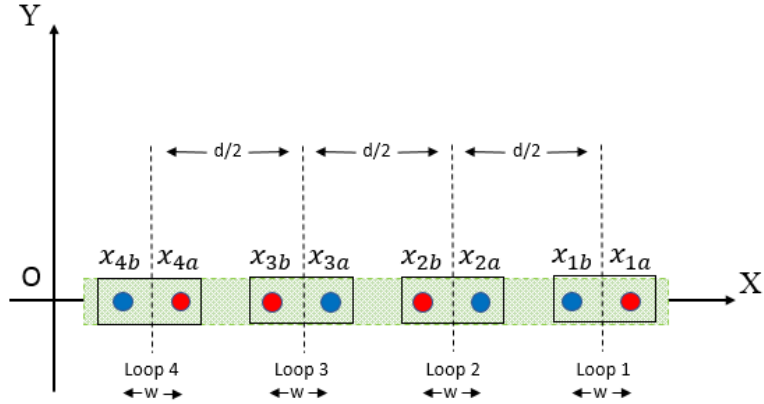

Figure 3: Cross-section of a single layer PCB with four loops each of one turn in DQB configuration. Used as an inductive pick-up coil, the PCB would rotate in the XY plane about the origin O. Dipole bucking is achieved by combining loops 1 and 3 in opposition, and Dipole-Quadrupole bucking by further combination of loops 1-3 with 4-2. The red and blue colors indicate the chirality of the loop from start (red) to end (blue).

Though bucked windings are designed to perfectly remove the main fields, the actual quality of bucking is affected by non-idealities. The extent of bucking can be characterized by defining the bucking ratio (BR) as flux amplitude of a particular component being bucked, for example quadrupole, in the un-bucked winding, divided by its residual flux amplitude in the bucked winding. Typical values for good probes can be in the hundreds or even thousands.

# 3. PCB design and manufacturing

The PCB rotating coil sensor can be manufactured by different methods, for example thick film, rigid printed circuit board, flex circuit board, or 3D printed coil. The technology choice is driven by constraints such as geometrical precision, the adopted PCB design rules [12], the number of layers, etc. The rigid PCB, though having a particularly complicated manufacturing process [11], is the most commercially available and reliable both for miniaturized boards, with copper trace width of 50 μm, copper thickness below 10 μm, trace-to-trace distance 50 μm, and long boards with length greater than 1 m intended for large magnets. The choice of board parameters may have trade-offs between the desired PCB coil sensor performance and the manufacturing limitations, in turn depending on the size and complexity of the PCB (Fig. 4). The details of optimizing those choices, however, will not be discussed here. The goal in each case is to have the ensemble of wires parallel to and planar with the rotation axis. Since the high placement accuracy of traces is in the etching of each layer plane and not in the stacking of layers, it is better not to rely on layer combinations to achieve bucking, but rather to have each layer buck the main fields independently (as in Section II) – the layer ensemble will then of course also buck these fields.

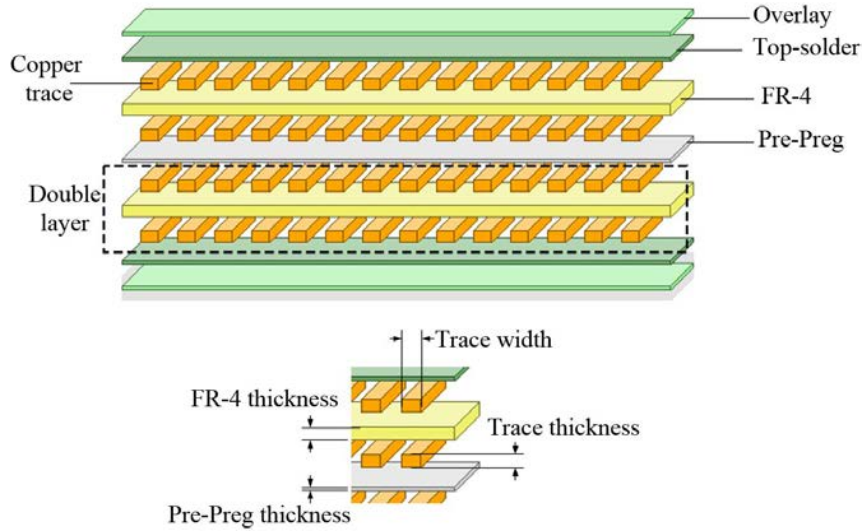

Figure 4: Section of a double-layer multi-layers PCB (pre-preg: pre-impregnated).

# 4. Rotating coil sensor calibration

In this section, the PCB calibration is discussed (i) with regards to calibration needs for good measurement of gradient (quadrupole) as main field, and (ii) with respect to calibration for higher-order harmonic content.

### 4.1) Calibration for gradient field measurement

Radial PCB-based sensors of the type discussed here have the advantage that if the Bucking Ratio (BR) is high (i.e. 500-1000) it is not necessary at some level to calibrate pick-up coil area and rotation radius for gradient measurement; the DB signal gives a radius-indpendent result which is accurate at about the 0.2% level.

This can be better explained by considering the measured gradient from the DB coil of Section 2 (Fig. 3). The gradient can be found from the flux (from (5)) as

$$C_2 = \frac{F_2^{track1} - F_2^{track3}}{K_2^{track1} - K_2^{track3}} = \frac{F_2^{track1} - F_2^{track3}}{\frac{L}{2R}[x_{1a}^2 - x_{1b}^2 - (x_{3a}^2 - x_{3b}^2)]} \quad . \tag{8}$$

With the definition of $\frac{C_2}{R} = g$ and $L = 1$

$$g_{calc} = \frac{2(F_2^{track1} - F_2^{track3})}{(x_{1a}^2 - x_{1b}^2 - (x_{3a}^2 - x_{3b}^2))} = \frac{2\Delta F}{dw} \quad . \tag{9}$$

If we now include that each trace has an asociated error, e.g. $\epsilon_{1a}$ for the trace at $x_{1a}$, then

$$g = \frac{2\Delta F}{((x_{1a} + \epsilon_{1a})^2 - (x_{1b} + \epsilon_{1b})^2 - (x_{3a} + \epsilon_{3a})^2 + (x_{1a} + \epsilon_{1a})^2 + (x_{3b} + \epsilon_{3b})^2)} \tag{10}$$

Expanding and using :
- $x_{1b} = x_{1a} - w$
- $x_{3a} = x_{1a} - d$
- $x_{3b} = x_{1a} - d - w$

$$g = \frac{2\Delta F}{(x_{1a}^2 - x_{1b}^2 - x_{3a}^2 + x_{3b}^2) + (2x_{1a}\epsilon_{1a} + \epsilon_{1a}^2 - 2(x_{1a} - w)\epsilon_{1b} - \epsilon_{1b}^2 - 2(x_{1a} - d)\epsilon_{3a} - \epsilon_{3a}^2 + 2(x_{1a} - d - w)\epsilon_{3b} + \epsilon_{3b}^2)} \tag{11}$$

Neglecting quantities that go as the error squared, $\epsilon^2$ :

$$g = \frac{2\Delta F}{2dw + 2x_{1a}\epsilon_{1a} - 2x_{1a}\epsilon_{1b} + 2w\epsilon_{1b} - 2x_{1a}\epsilon_{3a} + 2d\epsilon_{3a} + 2x_{1a}\epsilon_{3b} - 2d\epsilon_{3b} - 2w\epsilon_{3b}} \tag{12}$$

$$= \frac{2\Delta F}{2dw\left[\left(\frac{x_{1a}}{d}\frac{1}{DBR}\right) + \frac{1}{d}(\epsilon_{1b} - \epsilon_{3b}) + \frac{1}{w}(\epsilon_{3a} - \epsilon_{3b})\right] + 2dw} \tag{13}$$

where the Dipole Bucking Ratio (DBR) is:

$$DBR = \frac{w}{(\epsilon_{1a} - \epsilon_{1b}) - (\epsilon_{3a} - \epsilon_{3b})} \tag{14}$$

Finally, with the definition of the gradient without error from (9),

$$g \cong g_{calc} * \left[1 - \left(\frac{x_{1a}}{d}\frac{1}{DBR}\right) + \frac{1}{d}(\epsilon_{1b} - \epsilon_{3b}) + \frac{1}{w}(\epsilon_{3a} - \epsilon_{3b})\right] \quad (15)$$

The error in the gradient from Eqn. (15) is seen to be composed of three terms, which we identify as the relative size error between loop 1 and loop 3 (as contained in the DBR expression), the distance error between loop 1 and loop 3, and the absolute width error of the loops, respectively. In the case of multiple layers and traces, the error is the equivalent (average) error of all the traces of each loop section (discussed further below).

As an example of estimating the maximum expected error in gradient when using the DB winding, we take parameters from the probe used in this study (Section 5), $w$ = 5.5 mm, $x_{1a}$ = 34 mm, $d$ = 11.25 mm and apply these to the simple geometry of Fig. 3. Simulating random errors in trace placement of $\pm 2$ μm, the error in gradient is bounded by ~ 0.35% with a mean of 0.16% (Fig. 5). Note that the DBR given here is only affected by trace placement, and not other effects present in measurements, such as spurious dipole from noise, or relative tilt in planes of loop 1 and loop 3, which might make DBR appear lower but not affecting the gradient error. Let us also note that a typical complex PCB would average multiple turns, layers, and variation along the length to yield a net systematic error in trace placement, and might be substantially less than the 4 μm window for trace placement error used here.

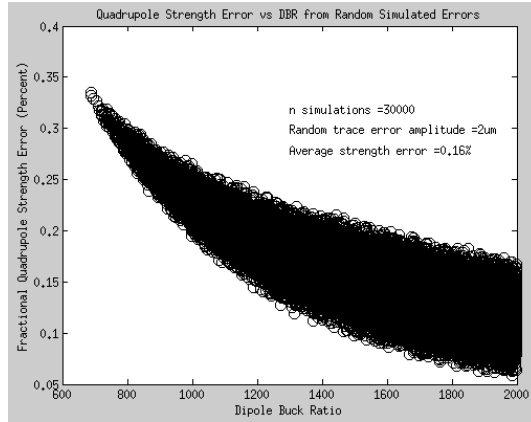

Figure 5: Quadrupole strength error vs DBR from simulated uncertainty. The simulation runs are 30000, the trace uncertainty is 2 μm, and the average strength error is 0.16%.

Two important results follow from this analysis: 1) given standard pcb manufacturing tolerances, the DB measured quadrupole gradient, eq. (9), will be accurate at the level of 0.1%-0.2% without any calibration; and 2) the measurement of this gradient (with its small error) is highly insensitive to horizontal (i.e. radial) positioning. To illustrate this second point, notice that the second and third terms of the error expression, eq. (15)), do not depend on the position of the windings at all, while the first error term depends linearly on $x_{1a}$ - so the change in the *error term* is the fractional error $\frac{\Delta x_{1a}}{x_{1a}}$. For example a (large) 10% change in $x_{1a}$ causes a change in error of 0.1 times the error term, so on the order of 0.01% for DBR of 1000.

Another important result is that the quadrupole strength measured by the DB winding does not depend on whether the PCB is vertically offset from the rotation axis (Fig. 7). To demonstrate this, we expand as before according to Eq. (2), except this time keeping the terms in y (assuming as before N=1, L=1):

$$K_2^{track1} - K_2^{track3} = \frac{1}{2} \{[(x_{1a}^2 - y_{1a}^2 + 2*x_{1a}*y_{1a}) - (x_{1b}^2 - y_{1b}^2 + 2*x_{1b}*y_{1b})] - [(x_{3a}^2 - y_{3a}^2 + 2*x_{3a}*y_{3a}) - (x_{3b}^2 - y_{3b}^2 + 2*x_{3b}*y_{3b})]\}, \quad (16)$$

Now setting $y_{1a} = y_{1b} = y_{3a} = y_{3b} = y$ (i.e. assuming the PCB is not bent or warped, but merely displaced vertically), we obtain

$$K_2^{track1} - K_2^{track3} = \frac{1}{2} \{[(x_{1a}^2 + 2*x_{1a}*y) - (x_{1b}^2 + 2*x_{1b}*y)] - [(x_{3a}^2 + 2*x_{3a}*y) - (x_{3b}^2 + 2*x_{3b}*y)]\}, \quad (17)$$

And since $x_{1a} - x_{3a} = x_{1b} - x_{3b} = d$, the remaining terms in y cancel and the DB winding sensitivity returns to the same expression as used in eq. (8), with no dependence on the offset in y. Since there is no change at all in DB sensitivity with vertical shift, we can treat both the amplitude and phase of the measured quadrupole as independent of both horizontal and transverse errors.

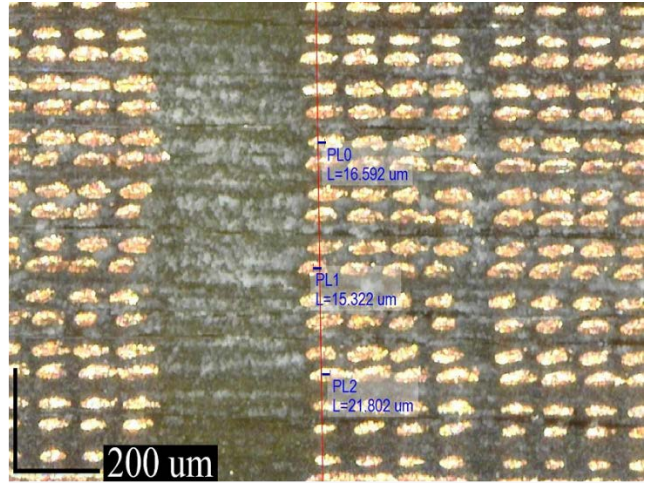

Figure 6: Microscope cross-section photo of PCB coil's control loops used to verify manufacturing errors (PLi: i-th distance measurement L of the trace actual center from the design reference (red line).

In the case of multiple turns for each loop in a single plane (Fig. 6), one can consider that as long as the other loops of the same layer also have the same multiple turns, that the bucking is preserved, yielding the same errors as discussed above. For multiple layers (also shown in Fig. 6), each layer has its own planar bucking, so even if layers do not align well, or if the thicknesses of the multiple layers differ, the bucking is still preserved. In terms of calculating an overall sensitivity, we keep in mind that for small errors in horizontal positioning (which may happen as layers are misaligned during stacking) the relative error in harmonics goes as

$$\frac{\Delta C_n}{C_n} = \frac{n\Delta r}{r} \tag{18}$$

and since this is linear in the radial offset error, $\Delta r$, the equivalent sensitivity of the multiple layers will be well represented by the average sensitivity calculated from the ensemble calibration of the next section.

Finally, it should be noted that the independence of the strength on both the radius and offset for the bucked winding has been shown here in the case of a quadrupole with dipole bucked winding, but also applies to the measurement of a higher order main field when the orders below the main field have been bucked, for example in the case of measurement of sextupole when the dipole and quadrupole fields have already been bucked.

### 4.2) Calibration of PCB position inside rotating shaft for harmonic fields determination

For the PCB rotating coil sensor mounted inside a shaft, we define a coordinate system such that the PCB lies along the horizontal axis. This gives the radial direction of the PCB with respect to the rotation axis (Fig. 7) and defines the coordinates of the PCB wires for sensitivity calculations. In this frame, the PCB can have vertical ($D_v$) and/or horizontal ($D_h$) displacement from nominal position as illustrated in Figs. 8a and 8b.

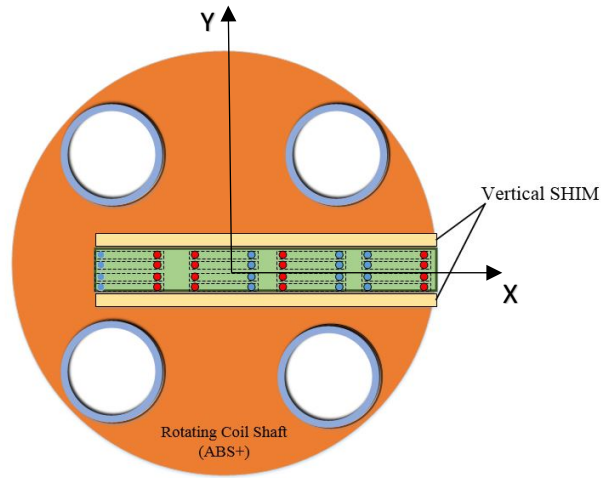

Figure 7 : Shaft section with PCB rotating coil sensor mounted in the design configuration (i.e. ideal mounting).

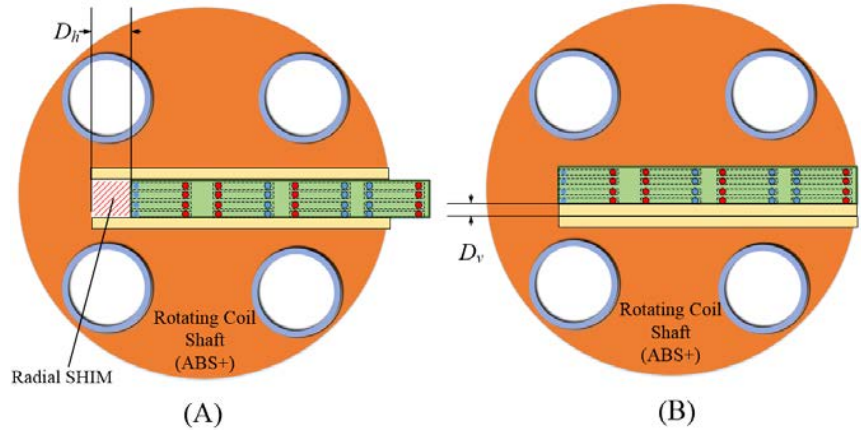

Figure 8 : Shaft section with PCB rotating coil sensor mounted with a horizontal shift $D_h$ (A) and with a vertical shift $D_v$ (B). The shift determined from calibration using the probe flux measurements is compared to the actual mechanical position.

Note that the error in PCB placement is specified fully by these two quantities, as any angular error in placement of the PCB within the shaft returns the same configuration as in Fig. 8 after rotation of the probe makes the PCB parallel (if not coincident) to the horizontal axis. Since the quadrupole field strength as determined by the DB winding is independent of the horizontal error ($D_h$) and vertical offset error ($D_v$) as shown in the previous section, and the quadrupole field as determined by the UB winding *is* sensitive to both these errors, the two PCB displacement errors can be determined by requiring that the DB quadrupole amplitude and phase match those of the unbucked (UB) winding. An example of the amplitude and phase dependence on the offsets is shown in Fig. 9, where the intersections of the windings gives the PCB displacement. The detailed analytical treatment follows.

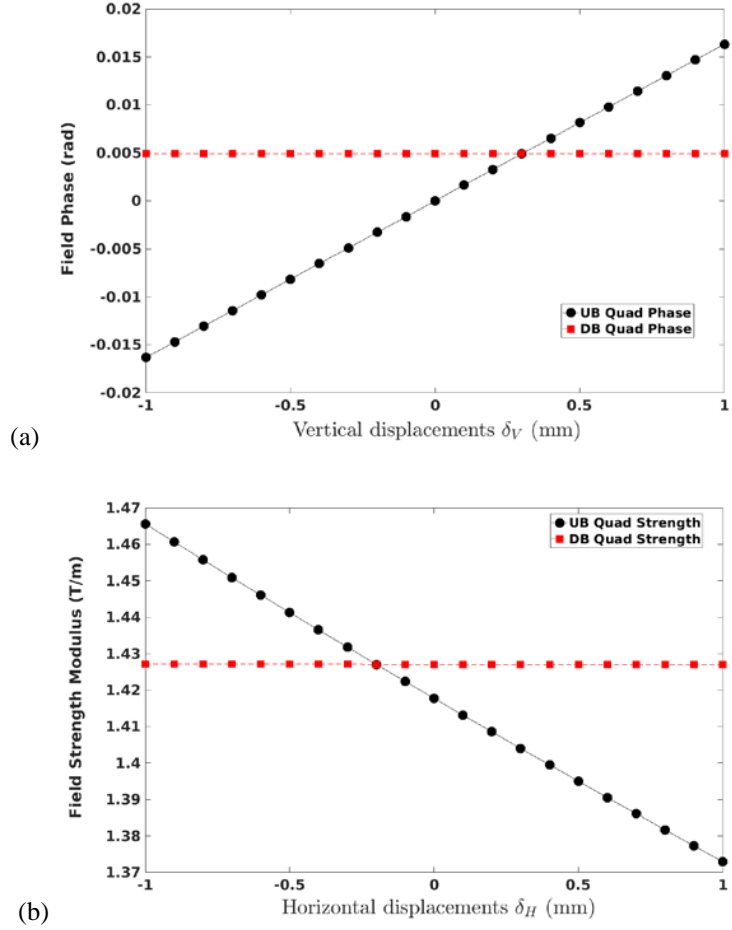

(a)

(b)

Figure 9: Field quadrupole phase (a) and strength (b) calculated for $K_2^{DB}$ (red) and $K_2^{UB}$ (black) relative to displacement increment of vertical $\delta_v$ (a) and horizontal $\delta_h$ (b) offsets.

## 4.3) Calibration of horizontal PCB position

Considering again the simple example of Figure 3 (and Eqn. 7b), the quadrupole strength measured by the DB winding of the coil (for N=1, L=1) is given by

$$C_2^{DB} = \frac{F_2^{DB}}{K_2^{DB}} = \frac{F_2^{Track1} - F_2^{Track3}}{K_2^{Track1} - K_2^{Track3}} = \frac{(F_2^{Track1} - F_2^{Track3}) * R}{d * w} \qquad (19)$$

whereas the quadrupole strength measured by the UB winding (loop 1) is

$$C_2^{UB} = \frac{F_2^{UB}}{K_2^{UB}} = \frac{F_2^{Track1}}{K_2^{Track1}} \tag{20a}$$

where the sensitivity is

$$K_2^{Track1} = \frac{1}{2R}[x_{1a}^2 - x_{1b}^2 + i*2y*(x_{1a} - x_{1b})] \tag{20b}$$

$$= \frac{w}{R}(\bar{r} + i*y) \tag{20c}$$

with y being the offset of the PCB plane from the radius, and $\frac{(x_{1a}+x_{1b})}{2} = \bar{r}$ the average radius of the UB winding. Unlike the DB winding, it is clear from $K_2^{Track1}$ that the UB winding is affected by both the evaluation of field amplitude (both by the $\bar{r}$ term and the imaginary term) as well as phase (since sensitivity is no longer pure real). Since the quadrupole field measured by both windings is the same, we set

$$C_2^{UB} = C_2^{DB} \tag{21}$$

and, combining (19) and (20a,c), the unknown quantities $\bar{r}$ and $y$ could be determined from evaluation of the real and imaginary parts of

$$\bar{r} + i*y = \frac{F_2^{Track1} d}{(F_2^{Track1} - F_2^{Track3})} \tag{22}$$

This is, of course, equivalent to requiring amplitude and phase of the UB and DB windings to be equal. Note that since typically $\bar{r} \gg y$ (usually $y$ has nominal value zero), the field amplitude, $|C_2^{UB}|$, depends predominantly on $\bar{r}$, and the phase on $y$.

In general, a solution can be found for any complex probe of many turns and layers. From Eqs. (19) and (20a), we can express the equivalence of the field measured with the DB and UB coils as

$$\frac{F_2^{Track1} - F_2^{Track3}}{K_2^{Track1} - K_2^{Track3}} = \frac{F_2^{Track1}}{K_2^{Track1\_cal}} \tag{23}$$

where we have used the superscript '_cal' to indicate on the RHS that this equivalence is true for the calibrated position of the UB winding (this is not used on the LHS since, importantly, DB is insensitive to positional error). In the horizontal direction the calibrated sensitivity can be written as

$$K_2^{Track1\_cal} = K_2^{Track1} + \text{Re}\left[\frac{\partial K_2^{Track1}}{\partial x}\right] D_h \tag{24}$$

where the real part of the derivative is used since we are interested in finding where equal values of the measured fields become the same from changes in the PCB position along the real (horizontal) axis. Together these equations give

$$\frac{F_2^{Track1} - F_2^{Track3}}{K_2^{Track1} - K_2^{Track3}} = \frac{F_2^{Track1}}{\left(K_2^{Track1} + Re\left(\frac{\partial K_2^{Track1}}{\partial x}\right) * D_h\right)} \tag{25}$$

and the radius error (parallel to the plane of the board as shown in Fig. 7) is given by

$$D_h = Re\left[\frac{F_2^{Track1} * (K_2^{Track1} - K_2^{Track3}) - (F_2^{Track1} - F_2^{Track3}) * K_2^{Track1}}{(F_2^{Track1} - F_2^{Track3}) * Re\left(\frac{\partial K_2^{Track1}}{\partial x}\right)}\right] \tag{26}$$

$$D_h = Re\left[\frac{F_2^{UB} * K_2^{DB} - F_2^{DB} * K_2^{UB}}{F_2^{DB} * Re\left(\frac{\partial K_2^{UB}}{\partial x}\right)}\right] \tag{27}$$

The sensitivity derivative term, $\frac{\partial K_2^{UB}}{\partial x}$, comes from first evaluating $K_2^{Track1}$ (from Eq. (2)) using the nominal values $x_j$ and $y_j$, and then repeating the calculation using wire locations shifted by $x_j \pm \delta_h$ (not changing the $y_j$). The derivative is then

$$\frac{K_2^{Track1}|_{(x_j+\delta_h)} - K_2^{Track1}|_{(x_j-\delta_h)}}{\delta_h - (-\delta_h)} \tag{28}$$

For a quadrupole, this simple two point slope determination is exact with respect to real part, since $\frac{dK_2^{Track1}}{dx}$ is linear (e.g. as in (*19b*)). For a magnet with main field of order *m* (sensitivity $K_m^{Track1}$), the linear assumption is typically valid only if the calculation $\delta$'s are close to the PCB offset ($D_h$). A higher order fit could be used, extending to larger offset error range as needed. However, and more simply, the above linear approximation to the derivative can be used and then iterated upon – i.e. the $D_h$ and $D_v$ (see next section) are added to the original nominal $x_j$ and $y_j$ and the $D_h$ and $D_v$ determinations of Eqns. 29 and 32 repeated. This achieves convergence quickly and with high precision even for large offsets. The iteration also removes any coupling effects of real and imaginary parts in determining harmonics of order *m* (i.e. applying the $D_h$ that was determined can affect the imaginary part of the $K_m^{Track1}$ (and likewise $D_v$, the real part), which may be substantial when calibrating in higher order fields.

For a magnet with main field order *m*, the horizontal offset error is given by

$$D_h = Re\left\{\frac{F_m^{UB} * K_m^{aB} - F_m^{aB} * K_m^{UB}}{F_m^{aB} * Re\left(\frac{\partial K_m^{UB}}{\partial x}\right)}\right\} \tag{29}$$

where the 'aB' superscript indicates a winding which has no horizontal dependence in determining field *m* (presumably one that bucks 'all' relevant fields below *m*).

### 4.4) Calibration of vertical PCB position offset

Similar to the horizontal offset determination, the vertical offset can be found in general for a probe with multiple turns and layers from considering the imaginary part of Eqn. (23). In the vertical offset case we use

$$K_2^{Track1\_cal} = K_2^{Track1} + \text{Im}\left\{\frac{\partial K_2^{Track1}}{\partial y}\right\} D_v \qquad (30)$$

and find $\frac{dK_2^{Track1}}{dy}$ from recalculating $K_2^{Track1}$ first using the nominal wire position values $y_j$ and $x_j$ (or positions $y_j$ and $x_j + D_h$, if $D_h$ has already been determined), and then repeating the calculation for $y_j \pm \delta_v$ and determining slope. Following the treatment as for $D_h$ we then arrive at

$$D_v = \text{Im}\left\{\frac{F_2^{UB} * K_2^{DB} - F_2^{DB} * K_2^{UB}}{F_2^{DB} * \text{Im}\left(\frac{\partial K_2^{UB}}{\partial y}\right)}\right\} \qquad (31)$$

Or more generally,

$$D_v = \text{Im}\left\{\frac{F_m^{UB} * K_m^{aB} - F_m^{aB} * K_m^{UB}}{F_m^{aB} * \text{Im}\left(\frac{\partial K_m^{UB}}{\partial y}\right)}\right\} \qquad (32)$$

where again, iteration using $D_h$ and $D_v$ values obtained for both *(29)* and *(32)* should be perfomed as needed to obtain exact results.

### 4.5) Calibration discussion

To improve measured main field strength accuracy beyond the typical 0.1-0.2% of Section 4.1, the rotating coil should have an absolute field referencing such as can be achieved (for a field integral) with a Single Stretched Wire (SSW) [17][18]. Alternatively, the error terms in eq. (15) can be largely eliminated by using a calibrated microscope to determine the trace positions through suitable cross-sectioned *control loops* (Fig. 6); the $K_n$ used for the calibration of Sections 4.3 and 4.4 would then incorporate these accordingly. The assumption in this case is that the control loops (which could be sampled at both ends of the PCB) would represent the average cross-section along the PCB length.

In terms of understanding the PCB cross-section further, layer misalignment can in principle also be determined using only the calibration technique without need of a microscope. This is possible because each layer has its own bucking, and would involve reading out the UB and DB signals for an individual layer, and repeating the calibration through each layer in turn. The relative displacements between the layers (horizontal shift) and the actual distance between the layers loops (vertical shift) would then be known. The resulting Kn, reflecting the actual layer locations, could then be used when calibrating the $D_H$ and $D_v$ of the complete coil ensemble.

Note that the PCB probe calibration can be carried out *in situ*, since it requires only knowledge of the PCB and not of the magnetic field; neither does it require additional specialized equipment or tests. Moreover, the calibration can be performed dynamically and repeated at any moment during measurements or as a separate procedure. Calibration can be carried out equally well in magnets with main field of order quadrupole or higher as noted, as long as the PCB has a signal which bucks relevant fields below the main field order (e.g. a Dipole-Quadrupole-Bucked (DQB) signal in a sextupole).

The PCB calibration can be used to mitigate some concerns of the measurement environment, e.g. temperature effects. Furthermore, it may compensate for mechanical changes over time – unlike typical calibrations which may have limited duration because of metrological constraints or from undesired shocks during handling and mounting. Conversely, the PCB calibration can be used to track changes in coil position over time and test conditions. Also, if the probe is used on a magnet shorter than the probe, the calibration appropriate to the section of probe is determined and applied, rather some overall average over the entire probe length.

## 5. Experimental validation

### 5.1. Test setup

The coil calibration technique was validated experimentally using a PCB with 9 turns per loop, 10 layers, and length 1.02 m, having UB, DB, and DQB circuits as described in Section 2. Measurements were conducted at Fermilab.

The PCB is mounted on a support cylinder made with 3D printed parts of 'ABS+' plastic, and is assembled together with four full-length G-10 rods. The probe assembly is rotated by an external drive with encoder and slip-rings for continous uni-directional rotation. The test subject is a permanent magnet quadrupole (designated PQP003-12) of mechanical length 0.6 m; note that the probe extends out both ends in order to minimize sensitivity to probe positioning and allow comparison to integral strength measurements (Fig. 10).

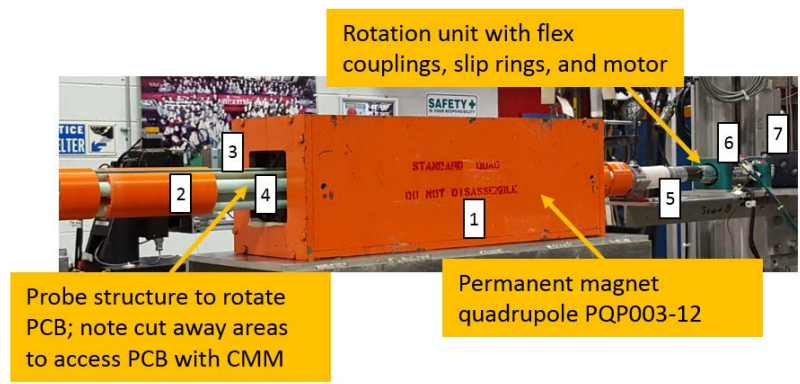

Figure 10: Test bench used for validating the calibration. 1) reference quadrupole, 2) probe body to support PCB, 3) PCB supported in probe structure, 4) underlying structure in area where body was cut away for CMM measurements, 5) flexible coupling, 6) slip-ring unit, 7) rotation motor

The probe body slot which holds the PCB is over-sized compared to the sensor thickness in order to allow for shifting the PCB vertically (i.e. $D_v$) by means of spacers (Fig. 11). Similarly, the rotating coil sensor can be shifted in the horizontal (radial) direction (i.e. $D_h$) by inserting spacers at the bottom of the slot. In both cases, the shims are of uniform, but not precise, dimension.

As a reference for the calibration validation, the position of the PCB sensor inside the shaft is measured by a Coordinate Measuring Machine (CMM). To measure the relative changes in horizontal and vertical displacement between PCB positions having various shims, the end portions of the support cylinder are milled to create reference surfaces directly on the probe itself. The vertical displacement is defined as the distance from exposed portions of the sides of the PCB to the midplane defined from measurements of the two sides of a radial slot (Fig. 12a). The horizontal displacement is calculated by measuring the PCB edge with respect to the top reference planes (Fig. 12b). Only the central portion of the CMM measurements, corresponding to the portion of the probe in the magnet, is used to determine the average shift of the probe caused by the shims. Gaps in the CMM data in later figures are regions where mechanical support of the PCB prevents measurement access.

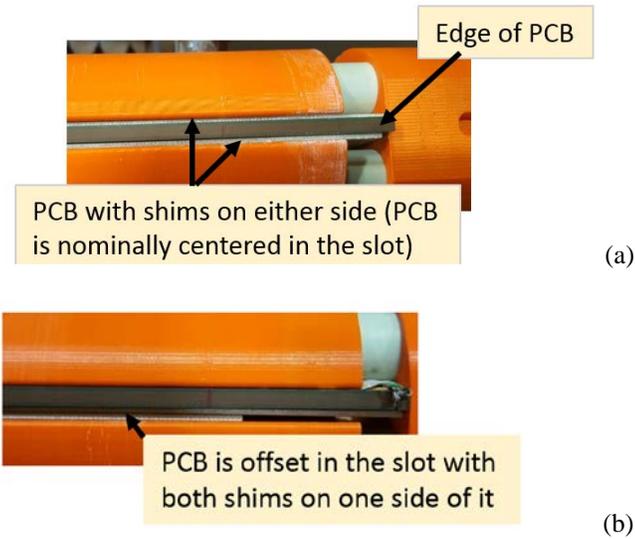

Figure 11: PCB sensor mounted in the 3D printed shaft (a) in central position and (b) shifted vertically with spacers (shims).

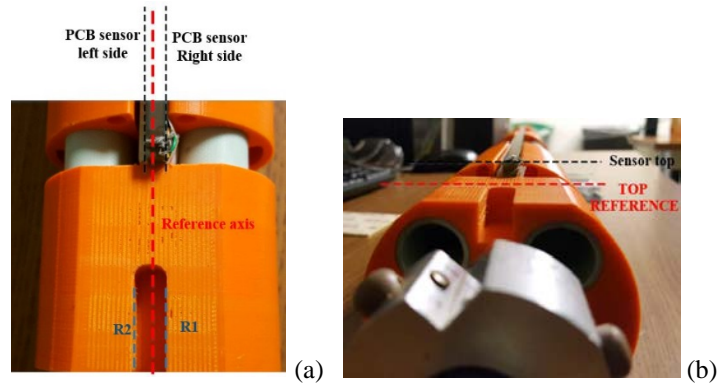

Figure 12: PCB rotating coil (a) radial slot, for vertical shift, and (b) top reference.

A LabVIEW™ program was implemented to acquire voltages during probe rotation using National Instruments™ 4462 24-bit ADCs. These were digitally integrated to determine flux at each of 1024 encoder angles. The calibration analyses of eqns. (27) and (31) were coded in Matlab™, and applied to each measurement rotation within the LabVIEW™ interface.

**5.2 Calibration results**

Determination of the PCB position with dynamic calibration was performed after CMM referencing measurements, both for nominal PCB position as well as with the PCB offset by horizontal or vertical shims. Within each set of the rotating coil measurements, the repeatability of the PCB position found by the calibration was better than 1 μm both horizontally and vertically.

For the vertical plane ($D_v$) referencing measurements, the stability between repeated CMM measurements two days apart, with the probe configuration undisturbed but re-mounted on the CMM bench, showed average difference below 7 μm. Radial changes ($D_H$), however, were observed as large as 40 μm after several days, depending on use and handling of the probe. To mitigate against this mechanical instability, rotating probe measurements for comparison were taken immediately after CMM data acquisition. Individual CMM measurements are shown in Fig. 13. To see how well the PCB shift determined from the rotating coil calibration compares to the shift measured by the CMM, an overlay of the original CMM data and CMM data including shim but shifted by the amount found from calibration, are shown in Fig. 14. If the calibration measures the same change as the CMM, then these two should perfectly overlay. Rather, it is observed that though the average overall profiles agree well, there are some local differences at perhaps the 50 μm level which indicate distortions of the probe and likely limit the comparison for this assembly. The average changes in PCB position in the absence and then presence of shims as measured by dynamic calibration and CMM are shown in Table 1. Close agreement is observed within the ~10 μm reproducibility of the measurements.

|  | Cal. shift (mm) | CMM shift (mm) |
|---|---|---|
| $D_v$ | 1.679 | 1.679 |
| $D_h$ | 6.298 | 6.304 |

Table 1: Comparison between horizontal, $D_h$, and vertical, $D_v$, shifts of the board caused by shimming as measured by the calibration technique described and by CMM measurements.

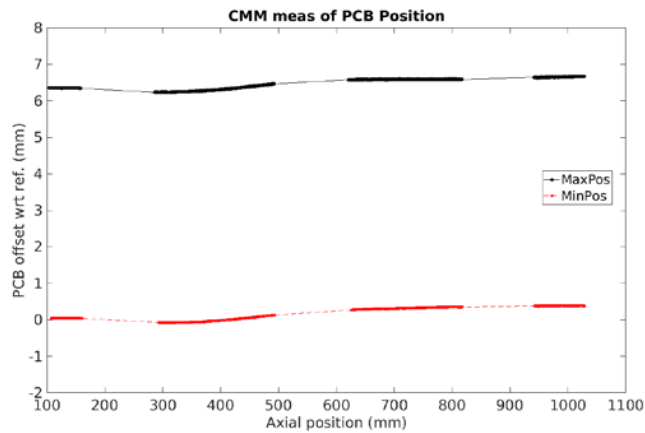

Figure 13: PCB edge measured with CMM along the coil longitudinal length (before and after the horizontal shift).

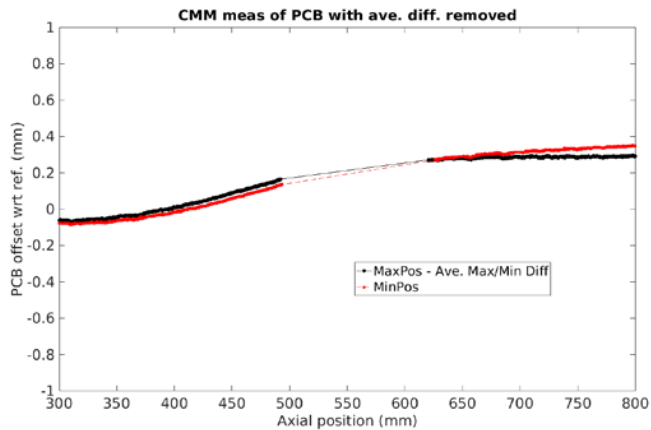

Figure 14: CMM data overlay of PCB sensor position with average measured difference, $D_H$ = 6.304mm removed. Local mechanical variations along the probe length are as large as 0.05mm. The average shift measured by the CMM and the calibration technique differed by 6 μm.

## 5.3 Field results

Quadrupole field strength of the PQP magnet as measured by the DB signal of the PCB probe was 1.5847 $\left[\frac{\text{T}-\text{m}}{\text{m}}\right]$ (s.d.m. ~0.00012). This compares very well to the strength measured by stretched-wire system of 1.5850 $\left[\frac{\text{T}-\text{m}}{\text{m}}\right]$. The effective dipole bucking achieved in the DB signal was ~560.

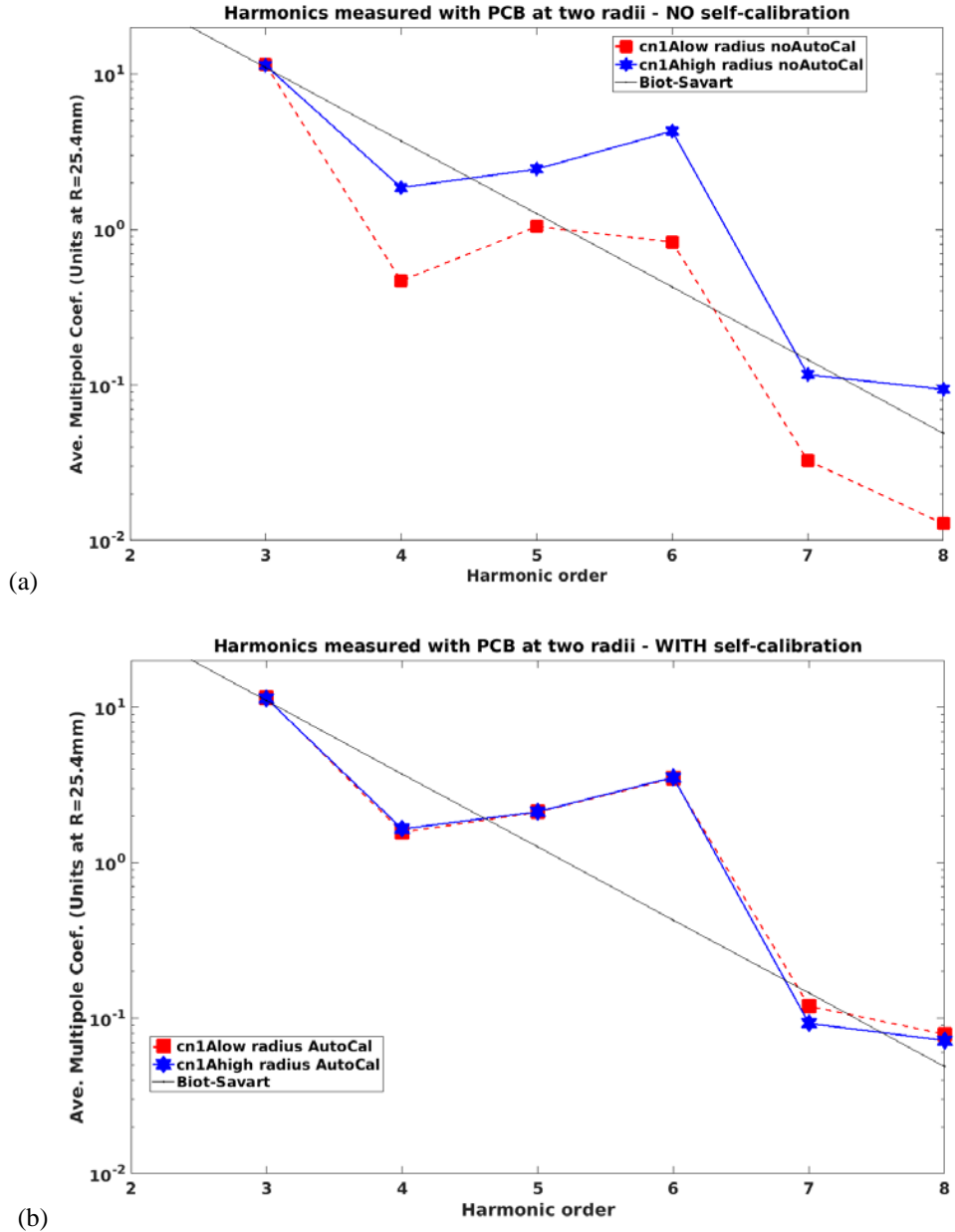

(a)

(b)

Figure 15: Harmonics amplitudes measured with PCB at two radial locations before (a) and after (b) dynamic calibration was applied. Note that the sextupole (n=3) values agree in both cases because these were measured with the probe DQB winding, which is radius independent for sextupole. The straight line approximates expected logarithmic fall-off with harmonic number.

As an example of the effectiveness of the calibration, the amplitude of multipole harmonics in 'units' of 1e-4 of main field at reference radius measured with the PCB at two unknown radial locations in the probe support cylinder (one as low as possible, the other as high as possible) are shown in Fig. 15. Data are presented both where no self-calibration is applied, using only some nominal values which were about 1.2 mm (smaller than the determined 'high' radius), and where the self-calibration has been turned on in the analysis code. Only results through n=8 are plotted since the 'low-radius' measurements were dominated by noise at the higher orders. Tables of the calibrated normal (bn) and skew (an) harmonics through n=10 are shown in Table 2. Very good harmonics agreement is observed when auto-calibration was applied, even though the difference in PCB radial positions in this case was very large.

|  | Low radius; after calibration | | | | High radius; after calibration | | | |
| --- | --- | --- | --- | --- | --- | --- | --- | --- |
| n | an | bn | s.d. an | s.d. bn | an | bn | s.d. an | s.d. bn |
| 2 | 0 | 10000 | 0 | 0 | 0 | 10000 | 0 | 0 |
| 3 | -8.35 | 13.66 | 0.01 | 0.02 | -8.27 | 13.51 | 0.02 | 0.02 |
| 4 | 2.24 | 1.32 | 0.03 | 0.05 | 2.4 | 1.3 | 0.02 | 0.02 |
| 5 | -3.05 | 2.79 | 0.03 | 0.03 | -3.02 | 2.82 | 0.02 | 0.01 |
| 6 | 1.32 | -7.91 | 0.06 | 0.08 | 1.08 | -8.07 | 0.01 | 0.03 |
| 7 | -0.29 | -0.15 | 0.04 | 0.07 | -0.23 | -0.12 | 0.02 | 0.01 |
| 8 | 0.27 | 0.09 | 0.24 | 0.2 | -0.17 | -0.15 | 0.04 | 0.01 |
| 9 | -0.31 | 0.53 | 0.32 | 0.13 | 0.01 | 0.05 | 0.03 | 0.02 |
| 10 | -0.13 | 0.66 | 0.68 | 0.73 | -0.02 | 1.74 | 0.05 | 0.02 |

Table 2: Harmonic normal and skew coefficients for the measurements of Figure 15. The amplitude shown in the Figure is defined as the square root of the sum of the squares of the bn and an.

## 6. Conclusion

A calibration technique has been developed for the fabrication of PCB-based rotating coils with high accuracy for the suppression of the dominant fields. The calibration determines the radial location and vertical shift of the PCB plane up to micron accuracy, and it is based only on the knowledge of the PCB itself and measured signals, not of the field or the mounting of the PCB in its rotating fixture. This is more than sufficient for ppm-level harmonics measurements, even for high orders, since vibration errors from *non-main field* harmonics begin to limit the accuracy of field results at that level. In terms of main field strength, even without calibration, PCB sensing coil geometry is expected to give accuracy of ~0.2% independent of the PCB radial location. The dependence on radial position is very weak, about 0.01% for a 10% change in radius, provided that the bucking ratio is the typical value of ~1000. The strength accuracy beyond that would need cross calibration to a reference, or microscope determination of the actual wire positions in the PCB cross-section based on 'control loops' at the PCB ends. If these exact wire positions in the cross-section were so determined, the actual cross-section could be used as the starting point for the PCB position calibration, in principle improving its accuracy further.

This calibration technique was subjected to experimental validation. This showed that the PCB offsets determined by the calibration matched physical measurements with Coordinate Measuring Machine (CMM) both for horizontal and vertical displacements to within the uncertainty of the mechanical measurements (10 $\mu m$), The PCB position calibration results were stable/repeatable at the 1 $\mu m$ level. The magnet integral strength determined during test was within 0.02% of the standard reference value. Furthermore, application of the calibration successfully determined correct harmonics from the measured data to better than ~0.1 units, even when the position of the PCB within its cylindrical support was not known at all *a priori*. The calibration technique was used *in situ* and applied dynamically during measurements, as could be done during any testing of magnets having main field of quadrupole or higher order.

## References

[1] J. T. Tanabe, "Iron Dominated Electromagnets: Design, Fabrication, Assembly and Measurements", World Scientific Publishing Company (May 2005), ISBN-10: 9812563814.

[2] S. Russenschuck, "Field Computation for Accelerator Magnets: Analytical and Numerical Methods for Electromagnetic Design and Optimization", Wiley (2011), ISBN: 978-3-527-40769-9..

[3] A. Jain, "Measurement and Alignment of Accelerator and Detector Magnets", CAS - CERN Accelerator School, Anacapri, Italy, 11 - 17 Apr 1997, pp.175-218 , ISBN: 9290831324.

[4] O. Dunkel, "Coil Manufacture, Assembly and Magnetic Calibration Facility for Warm and Cold Magnetic Measurements of LHC Superconducting Magnets at CERN", 14th International Magnetic Measurement Workshop, Contribution ID : 11, 26-29 September 2005, Geneva, Switzerland.

[5] M.Buzio, "Manufacturing and calibration of search coils", CAS on magnet, Bruges 2009, ISBN: 9789290833550.

[6] P. Arpaia, M. Buzio, G. Golluccio, F. Mateo, "In-situ calibration of rotating coil magnetic measurement systems a case study on Linac4 magnets at CERN", 17th Symposium IMEKO TC4, Kosice, Slovakia, September 8-10, 2010.

[7] P. Arpaia, M. Buzio, O. Kostner, S. Russenschuck, G. Severino, "Rotating-coil calibration in a reference quadrupole, considering roll-angle misalignment and higher-order harmonics", Measurement, Volume 87, June 2016, Pages 74-82.

[8] J. DiMarco, "Application of PCB and FDM technologies to magnetic measurement probe system development", IEEE Trans.Appl.Supercond. 23 (2013) no.3, 9000505.

[9] J. DiMarco. H. Glass, P. Schlabach, C. Sylvester, J.C. Tompkins, "Influence of Mechanical Vibrations on the Field Quality Measurements of LHC Interaction Region Quadrupole Magnets", IEEE Trans. Appl. Supercond., vol. 10, no. 1, march 2000.

[10] S. Turner, "Measurement and alignment of accelerator and detector magnets", CERN Accelerator School CAS, Anacapri, Italy, September 11-17, 1997, ISBN: 9290831324.

[11] L. Bottura, M. Buzio, P. Schnizer, N. Smirnov , "A tool for simulating rotating coil magnetometer", LHC Project Report 559, 17th International Conference on Magnet Technology (MT17), 24-28 September 2001, Geneva, Switzerland.

[12] "Design Rules and materials", http://www.we-online.com/

[14] P. Arpaia, M. Buzio, O. Dunkel, G. Severino, "Performance analysis of miniaturized PCB coils for small-aperture magnets qualification", IEEE Sensors Conference, Busan, Korea, 2015.

[15] J. DiMarco, "Application of PCB and FDM Technologies to Magnetic Measurement Probe Development",IMMW18, Upton, NY, June 2013.

[16] J. DiMarco, "Rotating Circuit Board Probes for Magnetic Measurements",IMMW15, Batavia, IL, August 2007.

[17] G. Deferne, M. Buzio, N. Smirnov, J. DiMarco "Results of magnetic measurements with Single Stretched Wire (SSW)",13th International Magnetic Measurement Workshop, Stanford, California, May 19-22, 2003.

[18] L. Walckiers, "Magnetic measurement with coils and wires", CAS- CERN Accelerator School : Specialised Course on Magnets, Bruges, Belgium, pp.357-385, 16 - 25 Jun 2009, ISBN: 9789290833550.